\documentclass[useAMS]{mn2e}
\usepackage{epsfig}
\usepackage{amsmath}
\usepackage[usenames,dvipsnames]{color}
\newcommand{\kms}{\ensuremath{\,\rm{km}\,\rm{s}^{-1}}}
\newcommand{\Msun}{\ensuremath{\rm{M}_\odot}} 
\newcommand{\TZO}{T\.ZO}
\newcommand{\TZOs}{T\.ZOs}

\title[HV 2112: an extrinsic S star in our Galaxy]{Large proper motion of the Thorne-\.Zytkow object candidate HV 2112 reveals its likely nature as foreground Galactic S-star}

\author[Maccarone \& De Mink]{Thomas J. Maccarone\\ Department of Physics, Box
  41051, Science Building, Texas Tech University, Lubbock TX, 
  79409-1051, USA\\email:thomas.maccarone@ttu.edu
\newauthor Selma E. de Mink\\ Astronomical Institute ``Anton Pannekoek'', University of Amsterdam, Postbus 94249, 1090 GE Amsterdam, The Netherlands\\email:S.E.deMink@uva.nl
}

\begin{document}
\def\ltsim{\mathrel{\rlap{\lower 3pt\hbox{$\sim$}}
        \raise 2.0pt\hbox{$<$}}}
\def\gtsim{\mathrel{\rlap{\lower 3pt\hbox{$\sim$}}
        \raise 2.0pt\hbox{$>$}}}

\date{}

\pagerange{\pageref{firstpage}--\pageref{lastpage}} \pubyear{}

\maketitle

\label{firstpage}

\begin{abstract}
Using the Southern Proper Motion (SPM) catalog, we show that the candidate Thorne-\.Zytkow object HV~2112 has a proper motion implying a space velocity of about 3000\kms if the object  is located at the distance of the Small Magellanic Cloud.  The proper motion is statistically different from that of the SMC at approximately $4\sigma$ in SPM, although the result can drop to about $3\sigma$ significance by including the UCAC4 data and considering systematic uncertainties in addition to the statistical ones.  Assuming the measurement is robust, this proper motion is sufficient to exclude its proposed membership of the Small Magellanic Cloud and to argue instead that it is likely to be a foreground star in the Milky Way halo. The smaller distance and therefore lower brightness argue against its proposed nature as a Thorne-\.Zytkow object (the hypothesized star-like object formed when a normal star and a neutron star merge) or a super-Asymptotic Giant Branch (AGB) star.  Instead we propose a binary scenario where this star is the companion of a former massive AGB star, which polluted the object with via its stellar wind, i.e. a special case of an extrinsic S star. Our new scenario solves two additional problems with the two existing scenarios for its nature as Thorne-\.Zytkow object or present-day super-AGB star. The puzzling high ratio of the strength of calcium to iron absorption lines is unexpected for SMC supergiants, but is fully consistent with the expectations for halo abundances.  Secondly, its strong variability can now be explained naturally as a manifestation of the Mira phenomenon. 
We discuss further observational tests that could distinguish between the foreground and SMC scenarios in advance of the improved proper motion measurements likely to come from Gaia.

\end{abstract}


\section{Introduction}

The nature  of the star HV~2112, located in the same part of the sky as the Small Magellanic Cloud (SMC), has recently become the center of a vivid debate. Levesque et al. (2014) showed that it exhibits anomalous spectral features suggesting enhanced abundances of molybdenum, rubidium, lithium, calcium and possibly also potassium when compared to other SMC supergiants. The high inferred bolometric luminosity and the enhanced abundances of the first three elements  lead to the intriguing suggestion by Levesque et al. (2014) that HV~2112 could be a massive ``Thorne-\.Zytkow Object'' (\TZO), the hypothesized red-giant like stellar objects with neutron stars inside their cores. 

\TZOs, first modeled by Thorne \& \.Zytkow (1975, 1977), may form as a result of unstable mass transfer in a massive X-ray binary after the neutron star is engulfed in the envelope of its companion star (Taam et al. 1978). Alternatively, these objects may form when the supernova kick of newly formed neutron star accidentally sends it towards and into its companion (Leonard et al. 1994). Whether \TZOs~actually form is a matter of debate.  Fryer et al (1996) and Chevallier (1993) argued that the neutron star would accrete enough material to collapse to a black hole before forming a stable \TZO.  Biehle (1991, 1994) and Cannon (1993) predict that massive TZOs, if stable, are powered by nuclear burning through the exotic interrupted rapid proton (irp)-process. Therefore they may be identified by enhanced abundances of rare proton-rich elements, including Mo and Rb. Also enhancements of Li have been predicted (Podsiadlowski et al. 1995).  The predictions triggered several (proposed) observing campaigns hoping to identify massive TZOs (van Paradijs et al. 1995, Kuchner et al. 2002, Levesque et al. 2014). So far HV~2112 is the most promising candidate, even though the abundances of Rb and Mo are not as high as expected and the enhancement of Ca and K was not predicted. 

Tout et al. (2014) discuss an alternative explanation where HV~2112 is a ``super asymptotic giant branch (AGB) star''. This is the late evolutionary phase of stars with initial mass around 8\Msun,  massive enough to ignite carbon, but still low enough to experience thermal pulses and dredge up episodes typical for AGB stars.  Super AGB stars have higher core masses than typical red giants, giving them luminosities as high as red supergiants and \TZOs in their early phases (Siess 2010). Similar to AGB stars they can, in principle, produce Mo and Rb through the slow neutron capture process (Karakas \& Lattanzio 2014) and Li through the Cameron \& Fowler (1971) mechanism.  Enhancement of Ca or K is, however, not predicted. 

In both scenarios HV~2112 is assumed to be member of the SMC. This view is supported by its projected location in the sky and the kinematic properties: the measured radial velocity (RV) is similar to that of the SMC (Neugent et al. 2010). However, as Levesque et al. (2014) state, the possibility that it may be a halo giant with a similar RV cannot be ruled out, but it would require a novel explanation for the peculiar spectral features.

In this paper, we report the detection of a very large proper motion based on data from the Southern Proper Motion survey (Girard et al. 2011). If confirmed, the proper motion excludes membership in the SMC, making HV~2112 a foreground Milky way star (likely a giant due to its pulsational properties). A location in the Milky Way would rule out both proposed scenarios, the \TZO-nature and sAGB-nature of HV~2112.

We propose a new scenario which involves pollution by the wind of a former massive AGB companion star.  In this manner, the enhancement of molybdenum, rubidium, and possibly lithium comes from the processes that would enrich a massive AGB star. By now the former companion has lost its entire envelope. Its remnant, most likely a massive white dwarf, is expected to be still around in a wide orbit.  HV~2112 in this scenario is the initially less massive secondary star. This scenario is similar to the scenario proposed by Iben \& Renzini (1983) to explain extrinsic S stars as well as  Ba and CH stars (McClure 1984). One of the very attractive features of this scenario (and any scenario were it is placed in the Galactic halo) is the natural explanation provided for the Ca (and K) features. These elements appear enhanced when compared to SMC giants, but they are normal for the Galactic halo where [Ca/Fe] enhancements are typical. We discuss various testable predictions of this scenario.

\section{Proper motion measurements and their implications}

The star HV~2112 has had its proper motion measured as part of the
Southern Proper Motion survey.  This is a systematic astrometric study
with a baseline of approximately four decades covering nearly the
entire southern sky down to approximately $V=17.5$ (Girard et al. 2011).  For HV~2112, the first epoch was obtained in 1972 and the second in 2007.  The star
shows a proper motion of 2.8$\pm$2.3 mas/year in right ascension and -9.8$\pm$2.3 mas/year in declination.  

While the star is strongly variable, there are two reasons to believe
that the proper motion measurements from SPM 4.0 are reliable.  First, in both
epochs, it was among the brightest stars in its general vicinity --
within two arcminutes, only one star is brighter in the second epoch V
band CCD data, and only two stars are brighter in the first epoch B
band photometry.  Additionally, within 2', there are only 8 stars brighter than $B=16$, so the probability of finding a star within 10" which is even 10\% as bright as HV~2112 is about $10^{-4}$.  Second, with the 35 year baseline, the total astrometric shift measured is about 0.35 arcseconds.  This shift is
sufficiently large that it would require extreme fine tuning to produce an astrometric shift due to the combination of blending and source variability without the second object being apparent in the images, and as shown, the probability of such a star being that nearby is of order $10^{-4}$.

The UCAC4 catalog (Zacharias et al. 2013) often gives an additional useful second epoch for proper motions to that from the SPM, although UCAC4 and SPM make use of the same first epoch data.  For this object, however, the second epoch from UCAC caught this variable star in a relatively faint state, so that the positional uncertainty is several times as large as in the SPM data.  The UCAC4 proper motion estimate is $1.8\pm2.9$ mas/yr in right ascension and $-3.3\pm2.7$ mas/yr in declination.  If one takes a weighted average of the two measurements, then proper motion is $2.4\pm1.8$ mas/year in right ascension and $-6.8\pm1.8$ mas/year in declination.  This differs from zero at roughly 3.8$\sigma$, and from the -1.1 mas/yr in declination of the SMC (Kallivayalil et al. 2013) by $3.2\sigma$.  The deviation from the proper motion of the SMC in right ascension, which is -0.8 mas/yr (Kallivayalil et al. 2013) is statistically insignificant, being about 1.7$\sigma$. Averaging SPM and UCAC using weights determined solely by their statistical uncertainties is a conservative approach, since any systematic uncertainties should be far more serious for the fainter UCAC4 epoch than for the SPM data, and the SPM measurement indicates a larger proper motion than does the UCAC4 measurement.

We have also checked for systematics in the astrometry of the field by taking the Hipparcos proper motion measurements for stars within 2 degrees of HV~2112 and cross-correlating them against the SPM 4.0 measurements.  We have found that there is no systematic offset at a
level of about 0.5 mas/year, and that the differences between the SPM and Hipparcos
measurements are comparable to those expected from the stated statistical errors.

We have test whether the internal errors for the SPM data are comparable to what is stated in the SPM catalogue.  Using the TOPCAT package, we have taken all the SPM stars within 1.5 degrees of HV~2112.  We have then filtered to include only stars whose 2MASS colours (Cutri et al. 2003) place them on one of the giant branches for the SMC and which have $V$ between 10 and 14 (i.e. a magnitude range for which we expect the errors to be similar to those for HV~2112), and then examined only those stars with proper motions of 5 masec/year or less.  We find that for these 199 stars, the standard deviation in the values of the proper motions in both dimensions is about 90\% of the mean values of the uncertainties on the proper motions from SPM 4.0.  The expectation, if all the stars had measurement uncertainties of the mean uncertainty, and there were no interloper stars, would be that the standard deviation would be about 75\% of the 1$\sigma$ uncertainty (the expectation is not exactly $1\sigma$ because the more extreme outliers will be eliminated by the 5 mas/year cut). It may then be that the measurement uncertainties are underestimated by about 20\%, and that the significance of the proper motion measurement relative to the SMC proper motion is at only the 3.0$\sigma$ level, but this gives us additional confidence that there are no major systematic uncertainties in SPM 4.0 that bias our results.

Any statistically significant measurement of a proper motion in data
with the precision of SPM 4.0 immediately provides a strong argument
against a location of the star in the Small Magellanic Cloud, because the statistical uncertainties on the proper motions are larger than the escape velocity from the SMC.  The proper motion in $\delta$ of 10 mas/year at a distance of 62.1 kpc (Graczyk et al 2014) yields a space velocity of 3100 km/sec, well
above the escape velocity of the SMC.  If one instead takes this to be
the velocity of a halo star in the Milky Way halo, and assumes a 150
km/sec speed (a reasonable value given the star's radial velocity of
157 km/sec -- Levesque et al. 2014), then it is likely to be located at a distance of about 3 kpc, although the statistical errors on the proper motion measurement allow for a wide range of possible distances.  Both the space velocity, and
the height, probably a few kpc below the Galactic Plane, given the
Galactic latitude of 45 degrees, suggest that this is a halo star
rather than a disk star in the Milky Way.
\begin{figure}
\centerline{\includegraphics[width=8 cm]{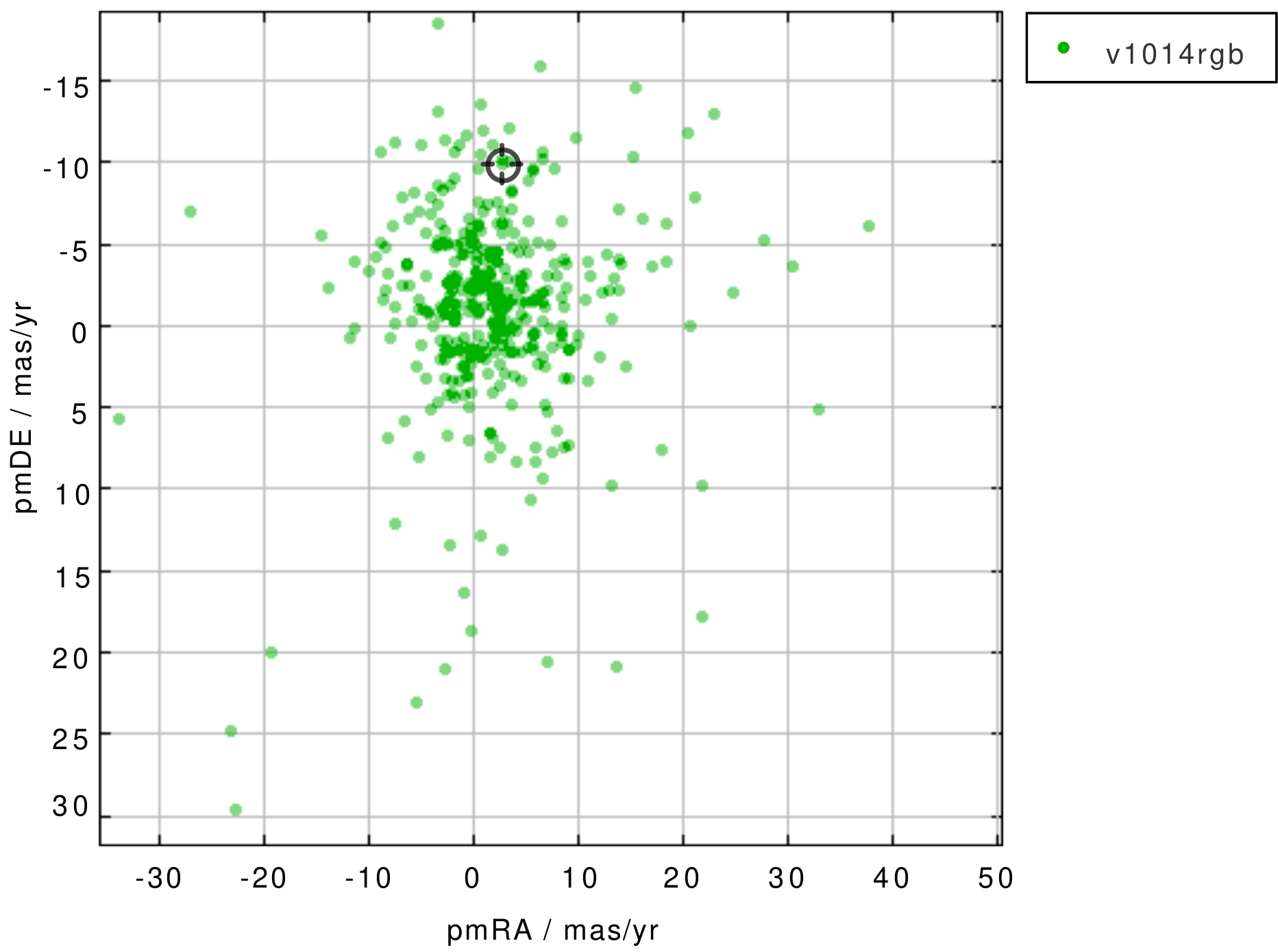}}
\caption{The proper motions of stars within 2 degrees of HV~2112, which have magnitudes from $V$=10 to $V=14$ and $J-K$ colors consistent with the giant branch for metal poor stars (from $J-K$=0.78 to 1.35).  HV~2112 has the black circle around it in the figure.  It can be seen to be an outlier in the proper motion diagram, with the main locus of points toward the center being the SMC stars, and the outliers being foreground stars.}\label{figpm}
\end{figure}

\begin{figure}
\centerline{\epsfig{figure=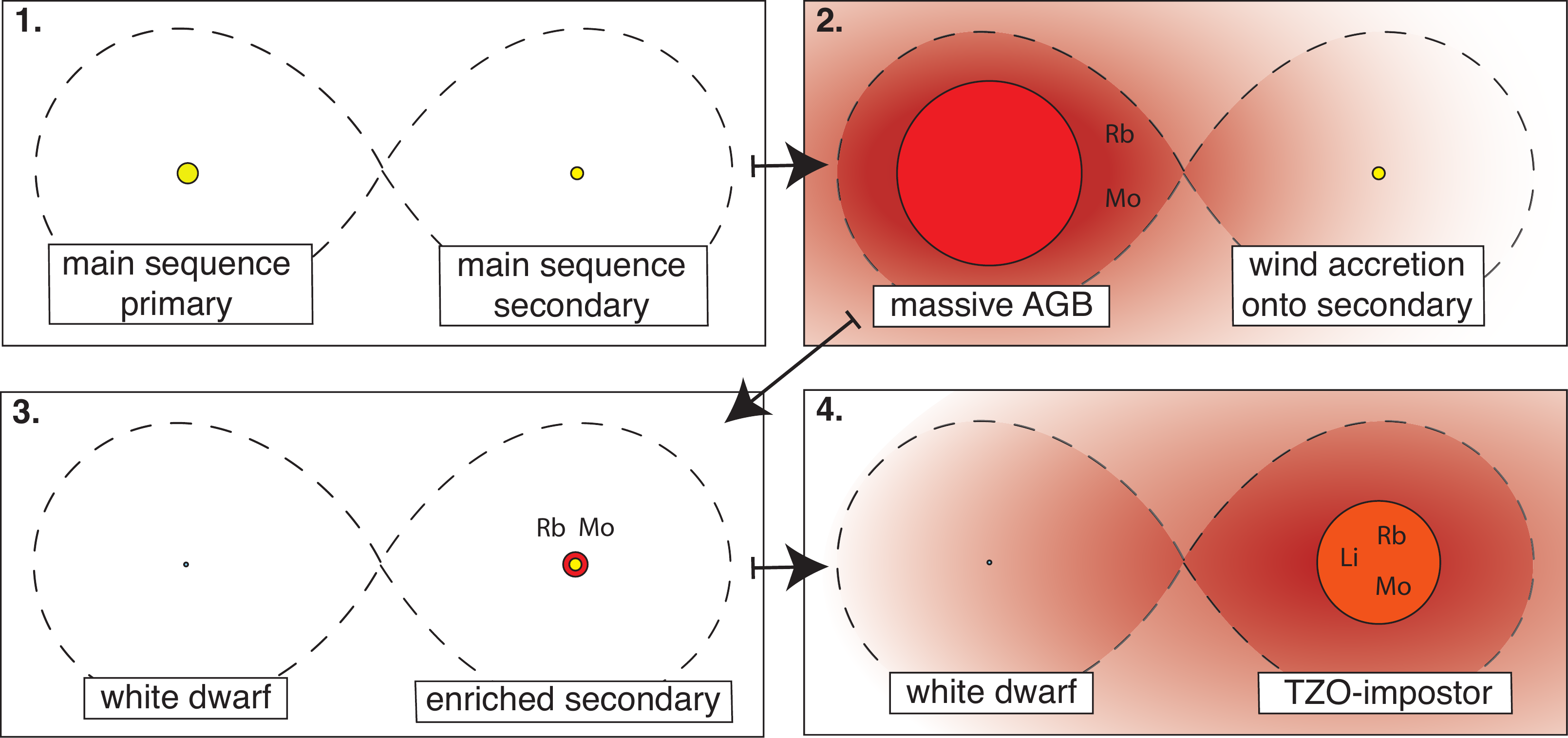, width=0.5\textwidth}}
\caption{{Cartoon of the alternative evolutionary scenario proposed involving a wide binary system located in the Galactic halo (1). The primary star evolves to become a massive AGB star and pollutes its low mass companion with its enriched wind (2), leaving behind a white dwarf after shedding its entire envelope (3). Eventually, the enriched secondary becomes an AGB star itself showing the, now diluted, abundance patterns of enrichment in addition to Li made in situ. The cartoon suggests that the white dwarf remnant of the primary is now accreting the wind of the ``TZO-impostor''. This is a possibility, but at present we have no strong evidence proving or disproving this hypothesis.
}}\label{fig:scenario}
\end{figure}

\section{A formation scenario for HV~2112}

At a distance of a few kpc, the luminosity of this object is
likely to be $\sim10^3 L_\odot$, rather than the $10^5 L_\odot$ it
would have in the SMC.  Furthermore, is has been classified as a Mira-like variable from photometry (Samus et al. 2012), and its emission lines in its bright phases are also consistent with Mira phenomenology (Levesque et al. 2014).  The hypothesis of a Mira star also fits well with the inferred luminosity for the closer distance.  These points were already noted in Levesque et al. (2014), but rejected on the basis of the then-likely association with the SMC.  Levesque et al. (2014) did briefly, but presciently, note that future kinematics observations could potentially place the object in the Milky Way, and that if they did, this would lead to a different interpretation than the Thorne-\.Zytkow object interpretation pursued in their paper.  Likewise, the super asymptotic giant branch (sAGB) star interpretation considered by Tout et al. (2014) also cannot explain the object if it is located in the Galactic foreground due to brightness considerations.

On the other hand, the abundances may be explained by a scenario in which matter is transferred from a massive AGB primary star to a binary companion.  Eventually, the secondary will evolve off the main sequence
and become an AGB star itself, albeit of significantly lower mass,
allowing such an object still to exist in the old population in the halo.  These stars are
generally referred to as extrinsic S stars (e.g. Iben \& Renzini
1983), following an initial purely phenomenological classification of
S stars (Merrill 1922), which were later found, in a mnemonically
fortunate coincidence to be rich in s-process elements (Smith \& Lambert 1986).  The primary must be massive enough to produce substantial enhancements of molybdenum and rubidium.  This certainly is expected for super-AGB stars (Lau et al. 2011; Tout et al. 2014), but is likely to happen down to even 4 $M_\odot$ (Lugaro et al. 2003), and is clearly seen in some S stars (e.g. Lambert et al. 1995 show that Rb abundances increase with increasing s-process elemental abundances, while Allen \& Barbuy 2006 show an excess of molbydenum in most of the members of a large sample of barium stars\footnote{The barium stars are thought to be an earlier evolutionary state of the extrinsic S stars}.).   The extrinsic S stars form through the binary evolution channel above,
while the intrinsic S stars are formed by dredge-up of s-process
elements, and show enhancement of Tc.  The Tc signature indicates that
the material has been placed into the stellar atmosphere on a
timescale of order or less than the radioactive decay time of Tc.

The placement of the star in the Milky Way halo also helps to explain
the calcium abundance anomaly that could not be explained by the super-AGB star scenario discussed in Tout et al. (2014) nor by the TZO
scenario of Levesque et al. (2014).  Levesque et al. (2014) did not
have a calibrated estimate of the Ca abundance of the star, but did
show that the ratio of strengths of Ca lines to Fe lines was a factor
of about 3 stronger than that for red supergiants in the SMC.  This is
indicative of a [Ca/Fe] ratio about 3.  Kobayashi et al. (2006) shows
that halo stars in the solar neighborhood with [Fe/H]$<-1$ tend to
have [Ca/Fe] of about 0.4, which indicates a calcium to iron abundance
ratio about 2.5.  The most metal rich (and presumably youngest) SMC
giants have ratios of $\alpha$ element abundances to iron abundances
quite similar to solar (Mucciarelli 2014).  Relatively little discussion exists about the potassium excess in HV~2112, but it is of similar size to the Ca excess judging from the plots in Levesque et al. (2014), and this, also, can be explained well in terms of empirical halo abundances (Kobayashi et al. 2006).

Lithium is a fragile element and is easily destroyed at temperatures in excess of about $2.5x10^6$K. Even a modest depth of mixing can lead to efficient depletion (Stancliffe 2009). In principle, Lithium can be in the AGB phase of the primary star (Cameron \& Fowler 1971) and accreted onto the secondary. However, all Lithium originating from the primary has most likely been destroyed when HV~2112 evolved to become an AGB star and developed a deep convective envelope. The present day lithium abundance is more likely the result of production in situ, in HV~2112 itself. Low mass AGB stars can produce lithium as a result of thermal haline mixing (Stancliffe 2010). An alternative possibility is that the white dwarf left behind accretes enough material to produce a substantial nova rate. Evidence for lithium production has been found in the classical nova V339 (Tajitsu et al. 2015).  This could pollute the envelope of HV~2112. The model predictions remain uncertain, but lithium is not uncommon in symbiotic Mira's.  For example V407 Cyg shows a strong lithium excess (Tatarnikova et al. 2003). 

\section{The variability of the system and its implications}
Radial pulsations have periods which scale with the dynamical timescale of the star -- that is, $P\propto{\rho^{-1/2}}$, where $P$ is the period of oscillation, and $\rho$ is the density of the star.  Mira variables have pulsation period of about 140-700 days.  Their masses tend to be $\approx 1M_\odot$ and their radii tend to be about 300 $R_\odot$.  The typical variations in radius are a factor of about 1.5 (e.g. Mahler et al. 1997).  Because the mass of the Thorne-\.Zytkow object could be quite large, it is viable for such a star to have a pulsational period of about 600 days caused by the same pulsation mechanism as the classical Miras, even though its radius would be much larger than a classical Mira.

On the other hand, the radial velocity amplitude of the pulsations should be quite a bit larger if the object is at SMC distances than if it is a foreground Galactic Mira variable.  In essence, one can apply the Baade-Wesselink method (Baade 1926; Wesselink 1946) to estimate the radius and hence distance to the object.  At SMC distances, the radius of the star should be about 1500 $R_\odot$, so a variation of 40\% about the mean radius is a variation of 600 $R_\odot$.  The pulsations of Miras in general, and HV~2112 in particular are such that most of the variation takes place in a relatively small range of pulse phase.  If we assume that such a size variation takes place in 0.2 times the pulse period, or 120 days, then we expect a velocity of expansion of about 50 km/sec.  This speed is not large enough to be inherently problematic for either scenario discussed here, but does provide a means for estimating the stellar radius, and hence the distance to the star if a radial velocity monitoring campaign is undertaken. 

\subsection{Population estimate}

We can make a crude estimate of the number of extrinsic S stars in the Galaxy with a massive AGB progenitors as part of testing the plausibility of the scenario.  S stars make about 4\% of giants in the temperature range where M stars are normally found, and extrinsic S stars make about 1/3 of S stars (van Eck \& Jorissen 2000).   Extrinsic S stars are thus about 1\% of the number of stars that appear to be M giants in color space. The SMC spans about $3\times10^{-4}$ of the solid angle of the sky.  The Milky Way has $\sim10^{11}$ stars, of which about $10^9$ are in the halo, and about $10^7$ are halo M giants.  There should then be approximately 30 extrinsic S stars in the Milky Way halo projected against the SMC. If we further assume that only the extrinsic S stars whose progenitors were more massive than 4 $M_\odot$ can have the required abundances of rubidium and molybdenum, we must correct for the fact that only 11\% of the stars between 1 and 10 $M_\odot$ (i.e. the ones that produce AGB stars without undergoing supernovae) are above 4 $M_\odot$ (and hence are heavy enough for production of these elements).  We then expect a few extrinsic S stars with massive enough progenitors in the SMC field. Since we additionally require that they be in an AGB state themselves to explain the Mira variations and the Li production, the expected number falls by a factor of about 30 (Girardi et al. 2010), so our expected number of objects is about 0.1.  The scenario we propose thus quite plausibly places a source of the type we invoke in a location projected against the SMC, while also not placing so many such objects projected against the SMC that we would have expected to have found many of them already.

\section{Additional testable predictions of the scenario}

The scenario makes several additional testable predictions.  First and foremost, the object should be in a wide binary system.  Orbital periods of extrinsic S stars are generally at least 600 days (Jorissen \& Mayor 1992).  The white dwarf in this system should likely be a ONeMg white dwarf, and hence should be of considerably higher mass that the S star.  Nonetheless, the motions should still be of
relatively low orbital velocity.  Assuming a total system mass of about 2
$M_\odot$ and an orbital period of about 2 years, the orbital velocity
of the optically bright star should be about 20 km/sec, while the
orbital velocity of the white dwarf should be about 10 km/sec. 

Additionally, the system may show some accretion by the white dwarf in
the system.  In Mira itself, for example, the white dwarf companion of
the optically bright star is seen to be accreting at about $10^{-10}
M_\odot$ per year (Sokoloski \& Bildsten 2010), but the orbital period
of Mira is about 500 years.  If, instead, this system has an orbital
period of only a few years, the accretion rate can be expected to be
much higher.  To first order, the mass accretion rate from a stellar
wind can be estimated to be:
\begin{equation}
\left(\frac{\dot{m}_{acc}}{\dot{m}_{w}}\right) = \left(\frac{v_{acc}}{{v}_{w}}\right)^4 \left(\frac{M_{acc}}{M_{acc}+M_{don}}\right)^2.
\end{equation}

Since the orbital speed of the donor star and the wind speeds of AGB
stars are of roughly the same order, and the accretor is heavier than
the donor, a substantial fraction of the total wind may be accreted (Bondi \& Hoyle 1944; Struck et al. 2004; Mohamed \& Podsiadlowski 2012). 

Given that the wind speed and orbital speed are not known in this
system, and the accretion rate depends on the ratio of these two
speeds to the fourth power, it is difficult to make a specific
prediction here.  If $\sim$ 1\% of the wind of a typical AGB star (i.e
1\% of $10^{-6} M_\odot$/yr) is accreted, however, then the system
would be expected to be bright in the ultraviolet due to strong
accretion.  The system is not detected in GALEX data, but does show a
blue excess in the spectrum of Levesque et al. (2014).  More targeted
searches for an accretion signature in this object are warranted, and
should be spread over the $\sim$600 day period of the optically bright star,
given that the accretion rate is likely to change with the pulsation
phase.  The system does not show X-ray emission, with a 90\% upper limit on its luminosity of about $10^{31} (d/6 {\rm kpc})^2$ in data taken on March 25, 2013 (V. Antoniou, private communication).  Relatively few cataclysmic variables have X-ray luminosities higher than this value, and those which do are frequent dwarf nova outbursters (e.g. Britt et al. 2015).  This limit, again, is not particularly constraining, as many symbiotic stars have X-ray luminosities below $10^{31}$ erg/sec.

\section{Acknowledgments}

We thank the Lorentz Center for having hosted the workshop on ``The Impact of Massive Binaries throughout the Universe'', where the ideas in this paper were first discussed.  We thank Stephen Justham,  Ben Davies, Lennart van Haaften and Onno Pols for useful discussions, Vallia Antoniou for sharing the X-ray upper limit in advance of publication, and Terry Girard for helpful and detailed discussions of possible systematic errors in the Southern Proper Motion survey data. SdM acknowledges support by the EU Horizon 2020 program through a Marie Sklodowska-Curie Reintegration Fellowship, H2020-MSCA-IF-2014, project id 661502.

\label{lastpage}

\end{document}